\documentclass[conference,a4paper]{IEEEtran}
    \newtheorem{theorem}{Theorem}
    \newtheorem{lemma}[theorem]{Lemma}
    
    \newtheorem{definition}[theorem]{Definition}

\usepackage[cmex10]{amsmath}
\usepackage{mathtools, amssymb, booktabs}
    \allowdisplaybreaks
    \def\>{\mathrel{-\mkern-2mu\triangleright}}
    \def\<{\mathrel{\triangleleft\mkern-2mu-}}
    \def\GG{{\mathbb G}}
    \def\Eo{E_0}
    \def\Er{E_\mathrm r}
    \def\Eaa{$\mathbb E$}
    \def\Ebb{\raisebox{-1ex}{\Eaa}}
    \def\Ecc{\scalebox{2}{\Ebb}}
    \def\Edd{\raisebox{1ex}{\Ecc}}
    \def\Eee{\Edd\rule[-5.5pt]{0pt}{0pt}}
    \def\bigE{\mathop{\Eee}\limits}
    \DeclareMathOperator\BSC{BSC}
    \DeclareMathOperator\dist{dist}

\usepackage{tikz-cd}
    \tikzset{every picture/.style={join=round,cap=round}}

\usepackage[noadjust]{cite}
\usepackage{hyperref}
    \hypersetup{
        colorlinks, allcolors=blue!75!green!80!black,
        pdfsubject={Information Theory (cs.IT)},
        pdfkeywords={
            polar code,
            finite blocklength,
            scaling exponent,
            channel degradation,
            dynamic kernel,
            hitting set,
        },
    }

\begin{document}

\advance\baselineskip 0pt plus.1pt minus.05pt
\advance\lineskip 0pt plus.1pt minus.05pt
\advance\parskip 0pt plus.2pt minus.1pt

\title{How Many Matrices Should I Prepare \\
    To Polarize Channels Optimally Fast?}

\author{%
    \IEEEauthorblockN{Hsin-Po Wang}%
    \IEEEauthorblockA{%
        University of California, Berkeley, CA, USA\\
        simple@berkeley.edu%
    }
    \and
    \IEEEauthorblockN{Venkatesan Guruswami}%
    \IEEEauthorblockA{%
        University of California, Berkeley, CA, USA\\
        venkatg@berkeley.edu%
    }
}

\maketitle

\begin{abstract}
    Polar codes that approach capacity at a near-optimal speed, namely
    with scaling exponents close to \(2\), have been shown possible for
    \(q\)-ary erasure channels (Pfister and Urbanke), the BEC (Fazeli,
    Hassani, Mondelli, and Vardy), all BMS channels (Guruswami,
    Riazanov, and Ye), and all DMCs (Wang and Duursma).   There is,
    nevertheless, a subtlety separating the last two papers from the
    first two, namely the usage of multiple dynamic kernels in the
    polarization process, which leads to increased complexity and fewer
    opportunities to hardware-accelerate.  This paper clarifies this
    subtlety, providing a trade-off between the number of kernels in the
    construction and the scaling exponent.  We show that the number of
    kernels can be bounded by \(O(\ell^{3/\mu-1})\) where \(\mu\) is the
    targeted scaling exponent and \(\ell\) is the kernel size. In
    particular, if one settles for scaling exponent approaching $3$, a
    single kernel suffices, and to approach the optimal scaling exponent
    of \(2\), about \(O(\sqrt{\ell})\) kernels suffice. 
\end{abstract}

\section{Introduction}

    For noisy-channel coding, both low-density parity-check code
    \cite{RiU08} and polar code \cite{Ari09} use the sum--product
    formulas to process information.  The product formula estimates $X_1
    + X_2$ given observations $Y_1$ and $Y_2$.  The sum formula
    estimates $X_2$ given
    (i) the difference $X_1 - X_2$ and
    (ii) the observations $Y_1$ and $Y_2$.
    Here, $(X_1, Y_1)$ and $(X_2, Y_2)$ are iid copies of some
    communication channel $W$.

    Since the publication of the original paper \cite{Ari09},
    researchers have found ways to improve polar code using
    generalizations of the sum--product formulas.  One generalization is
    to select an $\ell \times \ell$ invertible matrix $G$ and, for each
    $i \in [\ell]$, prepare a gadget that estimates the $i$th coordinate
    of $[X_1\, \dotsm\, X_\ell] \cdot G^{-1}$ (n.b.\ this is the product
    of a vector and a matrix) given
    (i) its first $i-1$ coordinates and
    (ii) the observations $Y_1, \dotsc, Y_\ell$.
    Here, $(X_1, Y_1)\,,\, \dotsc\,,\, (X_\ell, Y_\ell)$ are iid copies
    of some communication channel $W$.  Matrix $G$ is called a
    \emph{kernel}.  The sum--product formula is the special case $G
    \coloneqq [^1_1{}^0_1]$.

    Regarding how small the error probability can get when the code rate
    is fixed, the new gadgets generate a significant improvement: from
    $\exp(-\sqrt N)$ to $\exp(-N^{\beta(G)})$, where $N$ is the block
    length and $\beta(G)$ is some constant depending on $G$.  This
    $\beta(G)$ can be made arbitrarily close to $1$ if $G$ is carefully
    selected as $\ell \to \infty$ \cite{ArT09, KSU10, MoT14, HMT13}.
    And it is easy to show that $\beta(G)$ cannot take any value greater
    than $1$.

    Regarding the \emph{gap to capacity}, i.e., the capacity minus the
    code rate, while the error probability is fixed, the new gadgets
    have some advantages, but proving them is rather difficult.
    Generally speaking, channels can be classified into two types: those
    that are straightforward to work with and those that are not.  Over
    binary erasure channels, which belong to the former type, the gap to
    capacity can be improved from $N^{-1/3.627}$ to $N^{-1/\mu(G)}$,
    where $\mu(G)$ is called the \emph{scaling exponent} (of $G$) and
    can be estimated using the systematic method developed in
    \cite{FaV14, YFV19}.  There are also probabilistic arguments
    \cite{PfU19, FHM21} showing that, for a random $\ell \times \ell$
    kernel $\GG$,
    \[
        \mu(\GG) < 2 + O(\alpha), \qquad
        \alpha \coloneqq \frac {\ln(\ln\ell)} {\ln\ell}
    \]
    with a positive probability.  So there exists a series of kernels
    $G_2, G_3, G_4, \dotsc$ such that $\mu(G_\ell) \to 2$ as $\ell \to
    \infty$.  This limit, $2$, is the optimal scaling exponent a block
    code can have.

\begin{figure}
    \centering
    \begin{tikzpicture}
        \draw
            (0,0) node (W) {$W$}
            foreach \i in {1, 2, 3, 4}{
                (100-40*\i : 2) node (W\i) [scale=.6] {$W_G^{(\i)}$}
                (W) -- (W\i)
                foreach \j in {1, 2, 3, 4}{
                    (100-40*\i + 25-10*\j : 3) node (W\i\j) [scale=.36]
                    {$(W_G^{(\i)})_G^{(\j)}$} (W\i) -- (W\i\j)
                    foreach \k in {1, 2, 3, 4}{
                        (W\i\j) --
                        (100-40*\i + 25-10*\j + 6.25-2.5*\k : 3.5)
                    }
                }
            }
        ;
        \draw [xshift=5cm]
            (0,0) node (W) {$W$}
            foreach \i in {1, 2, 3, 4}{
                (100-40*\i:2) node (W\i) [scale=.6] {$W_{G(W)}^{(\i)}$}
                (W) -- (W\i)
                foreach \j in {1, 2, 3, 4}{
                    (100-40*\i + 25-10*\j : 3) node (W\i\j)
                    [scale=.36, inner sep=1]
                    {$(W_{G(W)}^{(\i)})_{G(W_{G(W)}^{(\i)})}^{(\j)}$}
                    (W\i) -- (W\i\j)
                    foreach \k in {1, 2, 3, 4}{
                        (W\i\j) --
                        (100-40*\i + 25-10*\j + 6.25-2.5*\k : 3.5)
                    }
                }
            }
        ;
    \end{tikzpicture}
    \caption{
        Left: one-kernel-for-all.  Select a fixed kernel $G$ and use it
        to polarize all channels.  Right: dynamic kernels.  A
        ``customized'' kernel $G(W)$ is selected for each channel $W$.
        Note that, as the code length increases, more channels are
        generated and naturally one would guess that more $G$'s are
        needed.  This work focuses on how to reduce the number of $G$'s
        needed.
    }                                                   \label{fig:tree}
\end{figure}
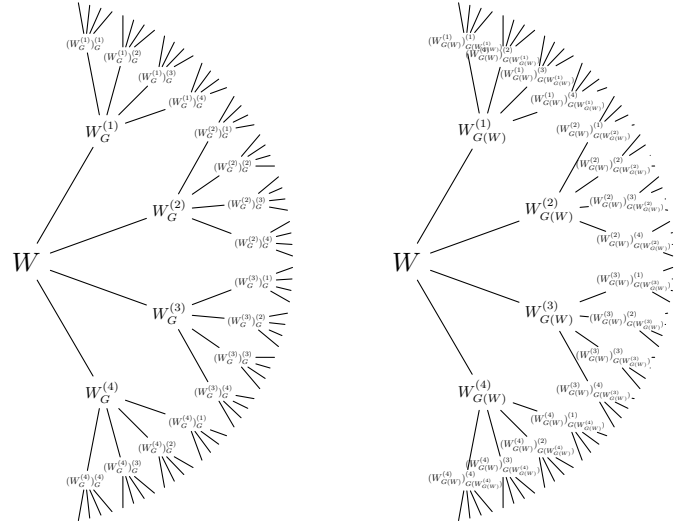

    If the underlying channels are less straightforward to work with,
    for instance binary symmetric memoryless (BMS) channels, then
    estimating $\mu(G)$ becomes a lot harder.  To demonstrate how hard
    it is, we remark that it is believed that $\mu([^1_1{}^0_1]) \approx
    4$ but the upper bounds we know are $\mu < 4.714$ \cite{MHU16} and,
    more recently, $\mu < 4.63$ \cite{WLV23}.  As for $\mu \to 2$, it is
    still possible, but now we need a collection of $G$'s, one for each
    BMS channel that occurs during the polarization process \cite{GRY22,
    WaD21}. See Fig.~\ref{fig:tree} for a visualization of this.  To put
    it another way, for a fixed matrix size $\ell$, BECs need $\ell$
    gadgets, but BMS channels need multiple $G$'s and each $G$ needs
    $\ell$ gadgets.  The more kernels we use, the more gadgets we need
    to prepare, and the less we can do about reusing or optimizing the
    gadgets.  In principle, the number of required kernels could grow
    linearly in the block length until it is capped by the number of all
    possible kernels, which is $\exp(-\Omega(\ell^2))$.

    The usage of channel-depending $G$'s is called \emph{dynamic
    kernels} \cite{YeB15}.  In this work, we aim to quantify how dynamic
    the kernels are.  Our contributions are twofold.  First, we show
    that channels close to each other can share the same $G$.  We call
    those a \emph{bundle} of channels.  We then count and find that
    there are
    \[ B = \exp( O( \ell^{1/\mu} ) ) \]
    bundles, where $\mu$ is the targeted scaling exponent.  Through
    this, we reduce the number of gadgets to $B \ell$.  As our second
    contribution, we use a counting argument to show that, if planned
    carefully, multiple bundles can share the same $G$.  This further
    reduces the number of required $G$'s to
    \[ m = O( \ell^ {3/\mu-1} ) \]
    and the number of required gadgets to $m \ell$.  Notice that $B$ and
    $m$ do not depend on the block length.  This is our main theorem.

    \begin{theorem} [main]
        At kernel size $\ell$ and for a targeted scaling exponent $\mu +
        O(\alpha)$, it suffices to prepare
        \[ m = O(\ell^{3/\mu-1}). \]
        binary matrices to polarize BMS channels.
    \end{theorem}

    The main theorem has two corollaries.  On the one hand, if a user
    aims for the optimal scaling exponent, $\mu \approx 2$, then the
    number of $G$'s grows linearly in $\sqrt\ell$; as a comparison, the
    total number of $\ell \times \ell$ matrices grows exponentially in
    $\ell^2$.  On the other hand, if a user relaxes the targeted scaling
    exponent to $\mu \approx 3$, then the number of necessary $G$'s
    stays constantly $1$; as a comparison, the only explicit matrix with
    a nontrivial upper bounds on $\mu(G)$ over BMS channels is the
    $\mu([^1_1{}^0_1]) < 4.63$ mentioned earlier.  There is a third,
    hidden corollary that if a user has a better upper bound on the
    number of bundles, then the number of matrices improves accordingly.
    For instance, BECs are totally ordered by degradation; hence, there
    are only $B = O(\ell^{1/\mu})$ bundles, and preparing $m = 1$ matrix
    is sufficient even when $\mu \to 2$; this recovers the results of
    \cite{PfU19, FHM21}.  Otherwise, when the channels are not totally
    ordered, we foresee an conflict between the optimal scaling exponent
    and the optimal number of matrices.

    The paper is organized as follows.
    In Section~\ref{sec:old}, we review the old proof of $\mu \to 2$.
    In Section~\ref{sec:trade}, we generate a trade-off between $\mu$
    and the probability that a random $G$ is good.
    In Section~\ref{sec:bundle}, we define a bundle of channels to be
    those that are close to each other.  We prove properties of bundles.
    In Section~\ref{sec:hit}, we compute the number of $G$'s
    given the number of bundles and the probability that a matrix is bad.

\section{How Scaling Exponent 2 Was Achieved}            \label{sec:old}

    While the scaling exponent $\mu$ of a code is defined to be the
    infimum of all positive numbers such that
    \[
        (\text{channel capacity} - \text{code rate})^\mu
        < \frac 1 {\text{block length}}
    \]
    for sufficiently large block lengths, existing methods that estimate
    $\mu$ can be summarized as
    \[
        \inf_h \sup_W \inf_G
            \frac {\sum_{i=1}^\ell h(W_G^{(i)})} {\ell h(W)}
        = \ell^{-1/\mu}.
    \]
    In this highly compressed notation enigma:
    \begin{itemize}
        \item $W_G^{(i)}$ is the virtual channel whose input is
            the $i$th coordinate of $[X_1\, \dotsm\, X_\ell] \cdot
            G^{-1}$ and the output is
            (i) the first $i-1$ coordinates alongside
            (ii) the observations $Y_1, \dotsc, Y_\ell$
            (here $(X_1, Y_1), \dotsc$  are iid copies of $W$);
        \item the inner infimum is taken over
            all $\ell \times \ell$ invertible matrices $G$;
        \item the supremum is taken over a set of channels $W$ that
            (a) contains the channel we want to transmit information over
                and
            (b) is closed under the operations $(\bullet)_G^{(i)}$;
        \item the outer infimum is taken over all functions $h$
            from the set of concerned channels to real numbers such that
            $h(\text{noiseless channel}) = h(\text{fully-noisy channel})
            = 0$ and $h > 0$ otherwise.
    \end{itemize}

    \cite{FHM21}, \cite{GRY22}, and \cite{WaD21} all used a fixed
    $h$-function: $h(W) \coloneqq H(W)^\alpha (1 - H(W))^\alpha$, where
    $\alpha \coloneqq \ln(\ln\ell) / \ln\ell$ and $H$ is the conditional
    entropy, so the outer infimum is instantiated.  The inner infimum is
    replaced by an expectation over a random matrix $\GG$ so we can
    apply Markov's inequality later.  In this work, we focus on BMS
    channels, so the channel capacity $I(W)$ is just $1 - H(W)$ and
    $\GG$ represent some binary matrix.

    With the previous paragraph and several other simplifications, the
    main problem is reduced to showing
    \begin{equation}
        \bigE_\GG \sum_{i=1}^j I(W_\GG^{(i)}) < \theta
        \quad\text{and}\quad
        \bigE_\GG \sum_{i=k+1}^\ell H(W_\GG^{(i)}) < \theta,
        \label{ine:E-CLT}
    \end{equation}
    where $\theta \coloneqq \exp(-\Omega(\ell^{2\alpha}))$
    and $j \coloneqq H(W)\ell - \ell^{1/2+\alpha}$
    and $k \coloneqq H(W)\ell + \ell^{1/2+\alpha}$.

    The meaning of the sum of $H$'s is as follows: Alice is sending
    information to Bob at rate $1 - k/\ell$.  The channel between Alice
    and Bob is $W$.  So Bob learns the entire message minus $\sum H$
    bits.  We want the amount of information Bob misses to be
    upper bounded by $\theta$.

    The meaning of the sum of $I$'s is as follows: Alice is sending
    information to Bob at rate $j/\ell$; Bob sees exactly what Alice
    sent (that is, the channel between Alice and Bob is noiseless).  Eve
    is an eavesdropper and the channel between Alice and Eve is $W$.  So
    Eve learns $\sum I$ bits of information.  We want the amount of
    information leaked to Eve to be upper bounded by $\theta$.

    To prove inequalities~\eqref{ine:E-CLT}, Gallager's $\Eo$ function
    and $\Er$ function are used.  The $\Eo$ function is the
    cumulant-generation function of $W$.  The $\Er$ function is the
    convex conjugate of $\Eo$, which is the Cramér function of $W$ as in
    the theory of large deviations.  This and Chang--Sahai's bound on
    the second moment of $W$ \cite{ChS07} imply
    \begin{align*}
        \Eo(t) & = I(W)t - \Theta(t^2), \\
        \Er(r) & = \Theta(I(W) - r)^2,
    \end{align*}
    for $t \approx 0$ and $r \approx I(W)$.  Since the code rate $r$ we
    care about is indeed close to the capacity, the amount of
    information Bob misses can be estimated using Gallager's error
    exponent argument as
    \begin{align*}
        \smash[b]{\bigE_\GG \sum_{i=k+1}^\ell H(W_\GG^{(i)})}
        & \approx \exp(- \ell \Er(1 - k/\ell)) \\
        & \approx \exp(- \ell \Er(I(W) - \ell^{-1/2+\alpha})) \\
        & \approx \exp(- \ell \Theta(\ell^{-1/2+\alpha})^2) \\
        & \approx \exp(- \Theta(\ell^{2\alpha})).
    \end{align*}
    Likewise, the amount of information leaked to Eve can be estimated
    using Hayashi's secrecy exponent argument as
    \begin{align*}
        \smash[b]{\bigE_\GG \sum_{i=1}^j I(W_\GG^{(i)})}
        & \approx \exp(- \ell \Er(1 - j/\ell)) \\
        & \approx \exp(- \ell \Er(I(W) + \ell^{-1/2+\alpha})) \\
        & \approx \exp(- \ell \Theta(\ell^{-1/2+\alpha})^2) \\
        & \approx \exp(- \Theta(\ell^{2\alpha})).
    \end{align*}
    After all approximation errors are accounted for, one obtains
    inequalities~\eqref{ine:E-CLT} contributing to the proof of $\mu \to
    2$.

\section{Negotiating with the Error Exponent}          \label{sec:trade}

    With the argument presented in the previous section, one infers that
    some random variables---namely the sum of $H$'s and the sum of
    $I$'s---are positive but their expectations are less than $\theta$.
    In past works, this is where one applies Markov's inequality with
    cutoff point $\theta$ and concludes that, with a positive
    probability,
    \begin{equation}
        \sum_{i=1}^j I(W_\GG^{(i)}) < \theta
        \quad\text{and}\quad
        \sum_{i=k+1}^\ell H(W_\GG^{(i)}) < \theta.
        \label{ine:CLT}
    \end{equation}
    Therefore, a suitable kernel exists for the channel $W$ in question.
    One then finds and uses a different kernel for each of the channels
    arising in the recursive construction.  In this paper, however, we
    want to modify this part of the argument as we want to reduce the
    number of distinct kernels that are employed.

    In this paper, we let $b$ be the probability that a random $\ell
    \times \ell$ matrix $\GG$ is ``bad.'' Bad means that one of
    inequalities in \eqref{ine:CLT} does not hold.  We want $b$ to be
    very small.  Intuitively speaking, this is because we want that any
    given bundle can be polarized by nearly all matrices so there is
    some luxury in matrix selection.

    Due to technical reasons, we do not want to relax the cutoff point
    $\theta$ in inequalities~\eqref{ine:CLT}.  So in order to reduce
    $b$, we relax $j$ and $k$ so that they are farther away from
    $H(W)\ell$.  This way, at the cost of a higher scaling exponent, we
    get a higher $\Er$, and hence we get a much smaller expectation.
    Then we can decompose this expectation into $\theta \cdot b$
    with the same cutoff point $\theta$ but a much lower $b$.

    More precisely, let $j$ and $k$ be $H(W)\ell - \ell^{1-1/\mu}$ and
    $H(W)\ell + \ell^{1-1/\mu}$, respectively, for some number $\mu > 2$
    that represents the relaxed target of scaling exponent.  Then
    \def\similar{
        \begin{smallmatrix}
            \text{similar}\\
            \text{calculation}
        \end{smallmatrix}
    }
    \[
        \bigE_\GG \sum_{i=1}^j I(W_\GG^{(i)})
        \approx \similar
        \approx \exp(- \Theta(\ell^{1-2/\mu+2\alpha})).
    \]
    Likewise,
    \[ 
        \bigE_\GG \sum_{i=k+1}^\ell H(W_\GG^{(i)})
        \approx \similar
        \approx \exp(- \Theta(\ell^{1-2/\mu+2\alpha})).
    \]
    Apply Markov's inequality with cutoff point $\theta$; the
    probability that $\sum H$ or $\sum I$ is greater than $\theta$ is
    less than
    \[
        \frac{\text{expectation}}{\theta}
        = \exp(-\Omega(\ell^{1-2/\mu})).
    \]
    It will be seen later that the formula gives interesting results
    when $2 \leq \mu \leq 3$.

    \begin{theorem}
        Suppose the targeted scaling exponent is $\mu$.  For any BMS
        channel $W$, a random kernel is bad with probability
        \[ b = \exp(-\Omega(\ell^{1-2/\mu})). \]
        Bad means that if we use this kernel to polarize $W$ in a code
        construction, we might not reach scaling exponent $\mu +
        O(\alpha)$.
    \end{theorem}

    Notice that, in the past works \cite{PfU19, FHM21, GRY22, WaD21},
    the authors only cared about existence of a good code instead of the
    probability that a random code is good.  This is because trading a
    low $\mu$ for a lower $b$ does not bring anything to the table.  But
    here, by the union bound, a random kernel is good for $1/b$
    channels.  So we do have an incentive to aim for a small $b$.  There
    are, however, many more than $1/b$ channels. So we need additional
    techniques in order to be able to use the same kernel for certain
    groups of channels.

\section{Bundling Similar Channels}                   \label{sec:bundle}

\begin{figure}
    \centering
    \begin{tikzpicture}
        \fill[black!10!white, t/.style={insert path={rectangle+(1,1)}}]
            (0,0)[t] (1,0)[t] (2,0)[t] (2,1)[t] (3,1)[t]
            (4,1)[t] (4,2)[t] (4,3)[t] (5,3)[t] (5,4)[t]
            (5,5)[t] (5,6)[t] (6,6)[t] (6,7)[t] (7,7)[t] ;
        \draw[gray, dotted] (0,0) grid (8,8);
        \draw[->] (0,0) -- (0,8) node[above]{cdf};
        \draw[->] (0,0) -- (8,0) node[below left]{$x \coloneqq h_2(p)$};
        \draw[brown] (0,0) -| (3,1) -| (5,3) -| (6,6)
            -| node [below right, inner sep=0] {$D(W)$} (7,7) -| (8,8);
        \draw[teal] (0,0) |- (2,1) |- (4,2) |- (5,4)
            |- node [above left,  inner sep=0] {$U(W)$} (6,7) |- (8,8);
        \draw[red!80!black, line width=1.2]
            (0,0) .. controls (7,) and (4,7) .. (8,8)
            node [pos=.73, left] {$W$};
        \draw[<->] (1,7) -- node[right]{$\ell^{-1/\mu}$} +(0,1);
        \draw[<->] (7,1) -- node[above]{$\ell^{-1/\mu}$} +(1,0);
    \end{tikzpicture}
    \caption{
        The quantification of a BMS channel is to ``select all tiles
        that intersect the red, thick path.'' For any BMS channel $W$,
        let $\omega$ be the measure on $[0, 1/2]$ as in the BMS
        decomposition of $W$.  The red path is the cdf of $\omega$ with
        a re-parametrization using $x \coloneqq h_2(p) \coloneqq -
        p\log_2(p) - (1-p)\log_2(1-p)$.  The upgradation $U(W)$,
        depicted in teal, corresponds to the path along the upper-left
        boundaries of the gray tiles.  The degradation $D(W)$, depicted
        in brown, corresponds to the path along the lower-right
        boundaries of the gray tiles.  The tile size is set to be
        $\ell^{-1/\mu}$.  Note that $H(W)$ is the area above-left to the
        red path.  Hence $H(D(W)) - H(U(W)) = 2\ell^{-1/\mu} -
        \ell^{-2/\mu}$.
    }                                                   \label{fig:tile}
\end{figure}
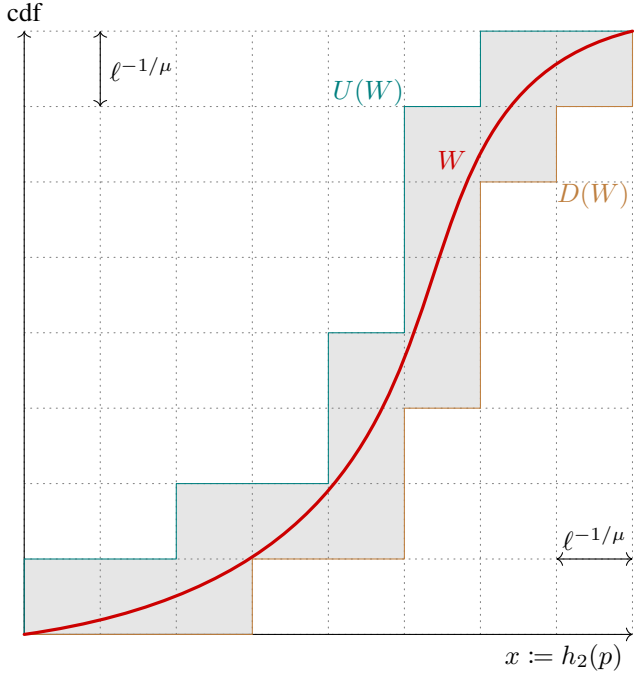

    In this section, we collect BMS channels that behave similarly into
    bundles.  We then show that channels in the same bundle can be
    polarized by the same $G$.  To measure similarity, we need the
    notion of degradation and upgradation.

    \begin{definition}[degradation]
        Let $\>$ denote the upgradation--degradation relation as in
        \cite{RiU08}.  When $U \> D$, we call $U$ an upgradation of $D$
        and $D$ a degradation of $U$.
    \end{definition}

    The following lemma connects degradation and polarization.

    \begin{lemma} [{\cite[Lemma~4.79]{RiU08}}]
        $U \> D$ implies $H(U) < H(D)$.  For any kernel $G$ and any
        index $i$, $U \> D$ also implies $U_G^{(i)} \> D_G^{(i)}$.
    \end{lemma}

    The overall strategy for classifying channels into bundles is that,
    for each channel $W$, we will define a degraded quantification
    $D(W)$ and an upgraded quantifications $U(W)$.  And then, channels
    that produce the same pair of quantifications go to the same bundle.
    This shares a similar spirit to \cite{PHT11, TaV13} but our
    quantification is not exactly the same as theirs.

    To quantize channels, the following notion is needed.
    
    \begin{definition}[BSC decomposition]
        For any BMS channel $W$, there exists a measure $\omega$ on $[0,
        1/2]$ such that
        \[ W = \int_0^{1/2} \BSC(p) \,\omega(dp). \]
    \end{definition}

    See Richardson--Urbanke \cite{RiU08} or Goela--Raginsky \cite{GoR20}
    for more formal treatment of BSC decomposition.  The meaning of the
    integral is that $W$ can be simulated by the following process.
    Pick a $P \in [0, 1/2]$ according to the measure $\omega$, and then
    feed the input into $\BSC(P)$ and reveal the channel output along
    with $P$.
    
    The following lemma connects BSC decomposition and degradation.
    Readers unfamiliar with degradation should note that the lemma is
    not an ``iff'' statement.  Cf.\ \cite[Theorem~4.74]{RiU08}.

    \begin{lemma}
        Let $U$ and $D$ be two BMS channels corresponding to measures
        $\upsilon$ and $\delta$, respectively.  If the cdf of $\upsilon$
        stays above the cdf of $\delta$, that is, if $\delta([0, p])
        \leq \upsilon([0, p])$ for all $p \in [0, 1/2]$, then $U \> D$.
    \end{lemma}

    Our construction of $D(W)$ and $U(W)$ is best described by
    Fig.~\ref{fig:tile}.  A formal description is also included for
    completeness: Suppose $n$ is the smallest integer greater than
    $\ell^{1/\mu}$.  For any measure $\omega$ on $[0, 1/2]$, let
    $\delta$ and $\upsilon$ be measures on $[0, 1/2]$ such that
    \begin{align*}
        \delta([0, h_2^{-1}(x)])
        & = \frac1n \Bigl\lfloor
            n\omega \Bigl(
                \Bigl[
                    0, h_2^{-1} \Bigl(
                        \frac {\lfloor nx\rfloor} n
                    \Bigr)
                \Bigr]
            \Bigr)
        \Bigr\rfloor, \\
        \upsilon([0, h_2^{-1}(x)])
        & = \frac1n \Bigl\lceil
            n\omega \Bigl(
                \Bigl[
                    0, h_2^{-1} \Bigl(
                        \frac {\lceil nx\rceil} n
                    \Bigr)
                \Bigr]
            \Bigr)
        \Bigr\rceil
    \end{align*}
    for all $x \in [0, 1]$, where $h_2^{-1}(x) \in [0, 1/2]$ is the
    inverse function of the binary entropy function.  Note that $x$ is
    the horizontal axis in Fig.~\ref{fig:tile}.  Now let
    \begin{align*}
        D(W) & \coloneqq \int_0^{1/2} \BSC(p) \,\delta(dp), \\
        U(W) & \coloneqq \int_0^{1/2} \BSC(p) \,\upsilon(dp)
    \end{align*}
    for $W$ corresponding to $\omega$.

    Two observations we make.  The first observation is that channels
    that pass the same set of tiles share the same $U(W)$ and $D(W)$.
    The second observation is that $H(D(W)) - H(U(W)) < 2/n$.

    \begin{lemma}
        $H(D(W)) - H(U(W)) < 2/n < 2\ell^{-1/\mu}$.
    \end{lemma}

    \begin{IEEEproof}
        $H(W) = \int_0^{1/2} h_2(p) \,\omega(dp)$ by the branching
        property of entropy.  So $H(W)$ is the area above the red curve
        in Fig.~\ref{fig:tile}.  Similarly, $H(D(W)) - H(U(W))$ is the
        number of shaded tiles times the area of each tile.  There are
        at most $2n - 1$ shaded tiles and each has area $1/n^2$.  this
        finishes the proof.
    \end{IEEEproof}

    \begin{definition} [pavement]
        Fix an $\ell > 0$ and a $\mu > 0$ (so $n$ is fixed).  A
        \emph{pavement} is a subset of $2n - 1$ edge-to-edge tiles that
        connects $(0, 0)$ to $(1, 1)$.
    \end{definition}

    \begin{definition} [bundle]
        A \emph{bundle} is a collection of BMS channels $W$ such that $U
        \> W \> D$, where $U$ and $D$ are the upper-left and lower-right
        boundaries, respectively, of a pavement.
    \end{definition}

    \begin{lemma}
        Every BMS channel is in at least one bundle.
    \end{lemma}

    By definition, the number of bundles is equal to the number of
    pavements, which is the number of north-east lattice paths from $(0,
    0)$ to $(n - 1, n - 1)$.  This number is $2(n - 1)$ choose $n - 1$,
    which is about $4^n$ or simply $\exp(O(n))$.

    \begin{theorem}
        Fix an $\ell > 0$ and a $\mu > 0$.  There are
        \[ B = \exp(O(\ell^{1/\mu})) \]
        bundles.
    \end{theorem}
    
    Remark: there is a subtlety we want to avoid when we are defining
    pavements and bundles: Consider what happens when the cdf of $W$
    passes a lattice point.  For instance, the red curve in
    Fig.~\ref{fig:tile} almost passes $(3/8, 1/8)$.  Then the cdf of
    $D(W)$ and the cdf of $U(W)$ will both pass the same lattice point.
    So the two tiles closest to that point are only vertex-to-vertex
    connected.  If we allow tiles in a pavement to be only
    vertex-to-vertex connected, then the number of pavements becomes the
    central Delannoy numbers, which is about $(3 + 2\sqrt2)^n$ instead
    of $4^n$.

    Now on why channels in the same bundle can be polarized by the same
    matrix, suppose $U \> W \> D$ and $G$ is a good kernel for $D$ and
    $U$.  Then
    \begin{align*}
        \sum_{i=1}^j I(W_G^{(i)})
        & \leq \sum_{i=1}^j I(U_G^{(i)}) < \theta, \\
        \sum_{i=k+1}^\ell H(W_G^{(i)})
        & \leq \sum_{i=k+1}^\ell H(D_G^{(i)}) < \theta,
    \end{align*}
    where $j \coloneqq H(D)\ell - \ell^{1-1/\mu+\alpha}$ and $k
    \coloneqq H(U)\ell + \ell^{1-1/\mu+\alpha}$.  Therefore, if $G$ is
    good for $D$ and $U$, it is good for all BMS channel bounded between
    $D$ and $U$.

    But since $H(D(W))$ and $H(U(W))$ are at most $2\ell^{-1/\mu}$ apart
    and $H(W)$ is sandwiched between those two numbers, this means that
    $k$ and $j$ are not too far away from $H(W)$:
    \begin{align*}
        j & > H(W)\ell - 3\ell^{1-1/\mu+\alpha}, \\
        k & < H(W)\ell + 3\ell^{1-1/\mu+\alpha}.
    \end{align*}
    Therefore, for the entire bundle bounded between $D$ and $U$, we
    only need to check if the random matrix $\GG$ suits $D$ and $U$.
    This means that, for a fixed bundle, $\GG$ is bad for channels in it
    with probability less than $b = \exp(-\Omega(\ell^{1-2/\mu}))$.

    \begin{theorem}
        For each bundle, a random matrix $\GG$ is bad with probability
        less than
        \[ b = \exp(-\Omega(\ell^{1-2/\mu})). \]
        Bad means that using this kernel whenever we see a channel in
        this bundle in a code construction will lead to a scaling
        exponent greater than $\mu + O(\alpha)$.
    \end{theorem}

    It is not hard to imagine that, if $b$ gets smaller, each bundle
    will accept more and more kernels.  So there is a higher chance that
    a kernel can serve multiple bundles.  The next section quantifies
    this statement.

\section{Hitting-Set Problem}                            \label{sec:hit}

    Suppose there are $B$ bundles and $b$ is the probability that a
    random $\GG$ is bad for a given bundle.  We claim that the number of
    necessary $G$'s is less than
    \[ m \approx -\log_b(B). \]

    To explain why, we use a straightforward greedy algorithm.  Imagine
    a bipartite graph where the left part consists of the bundles
    and the right part consists of the kernels.  A bundle is connected
    to a kernel if the kernel is bad for this bundle.  For a fixed
    bundle, $b$ is the probability that a random kernel is bad.  Hence
    the relative degrees of bundles are $b$.  By the Fubini principle,
    the relative degrees of the kernels have average $b$.  This means
    that there is some kernel who is bad for at most $b$ fraction of
    kernels.  Call this kernel $G_1$ and put it in the pocket.

    Now that all but $bB$ bundles are sufficiently polarized by $G_1$,
    it suffices to select more kernels for the remaining $bB$ bundles.
    We apply the same counting argument again to those $bB$ bundles, and
    derive that there is another kernel, called $G_2$, that is good for
    all but $b$ fraction of these $bG$ bundles.  Hence it remains
    to select more kernels for the remaining $b^2B$ bundles.  Repeating
    this, we see that for every extra kernel we select, the number of
    unsatisfied bundles reduces by $b$-fold.  We then conclude that it
    takes
    \[ m \leq -\log_b(B) + 2 \]
    kernels to reduce the number of unsatisfied bundles to $< 1$, 
    which is $0$ as it is naturally an integer.

    Plug-in $b = \exp(-\Omega(\ell^{1-2/\mu}))$ and $B =
    \exp(O(\ell^{1/\mu}))$.  The number of matrices needed is
    \[
        \frac{O(\ell^{1/\mu})}{\Omega(\ell^{1-2/\mu})}
        = O(\ell^{3/\mu-1}).
    \]
    This proves our main theorem.

    \begin{theorem} [main]
        At kernel size $\ell$ and for a targeted scaling exponent $\mu +
        O(\alpha)$, it suffices to prepare
        \[ m = O(\ell^{3/\mu-1}). \]
        binary matrices to polarize BMS channels.
    \end{theorem}
    
    Note that, if $\mu > 3$, $\ell^{3/\mu - 1}$ goes to zero as $\ell$
    goes to infinity.  Thus, at some point, it will eliminate the big
    $O$ constant and indicates that only $1$ (one) kernel is needed.

\section{Discussion: BMS Channel Geometry}

    Alongside the main theorem, one way to paraphrase the contribution
    of this paper is that we characterize the geometry of the space of
    all BMS channels.

    Define a distance between two BMS channels $W$ and $W'$ as
    \[ \dist(W, W') \coloneqq \inf_{D,U} H(D) - H(U), \]
    where the infimum is taken over all $D$ and $U$ such that $U \> W \>
    D$ and $U \> W' \> D$.  (Note that this is not a distance in a
    strict sense as we do not check the triangle inequality.)

    The concept of bundles in this paper is then analogous to epsilon
    open balls under this distance.  In Section~\ref{sec:bundle}, we see
    that the number of bundles is about $4^n$ when the diameter is about
    $2/n$.

    Usually, when a metric space has a finite Hausdorff dimension, the
    number of open balls should be inverse-polynomial in the radius of
    the balls.  Here, however, $4^n$ grows too fast, which suggests that
    the space of BMS channels behave as if it has infinite dimension
    under this distance.

\section{Conclusions}

    In this paper, we address a subtle distinction between \cite{PfU19,
    FHM21} and \cite{GRY22, WaD21}: the latter use dynamic kernels,
    which incur some complexity penalty.  We analyze how dynamic the
    kernels are, giving an upper bound on the number of matrices in
    terms of $\ell$ (the matrix size) and $\mu$ (the targeted scaling
    exponent).  Two special cases of our theorem are:
    (a) if $\mu \to 2$,
        then about $\sqrt\ell$ matrices are needed for each $\ell$;
    (b) if $\mu \to 3$,
        then one matrix is needed for each sufficiently large $\ell$.

    One might have evidence or choose to believe that the number of
    bundles is overestimated.  For instance, if one is dealing with
    BECs, then our theorem allows $\mu \to 2$ and $m = 1$ at the same
    time, which explains the usage of single kernel in \cite{PfU19,
    FHM21}.  Or, one might believe that all we need to polarize are
    additive white Gaussian noise (AWGN) channels.  Since AWGN channels
    are totally ordered under degradation, a single kernel and $\mu \to
    2$ can coexist just like the BEC case.

\section{Acknowledgment}

    This work was supported in part by a Simons Investigator award and
    NSF grant CCF-2210823.

\tracingpages1
\IEEEtriggeratref{10}
\IEEEtriggercmd{\enlargethispage{\dimexpr 273pt - 672pt}\pagebreak}
\bibliographystyle{IEEEtran}
\bibliography{DegradeWide-25.bib}

\end{document}